\documentclass[slac_one]{revtex4}
\usepackage{graphicx}
\usepackage{fancyhdr}
\begin{document}

\title{{\small{2005 International Linear Collider Workshop - Stanford,
U.S.A.}}\\ 
\vspace{12pt}
LCDG4 and DigiSim -- Simulation activities at NICADD/NIU
} 

%

\author{D.~Beznosko, G.~Blazey, D.~Chakraborty, A.~Dyshkant,
K.~Francis, D.~Kubik, J.G.R.~Lima, J.~McCormick, R.~McIntosh,
V.~Rykalin, V.~Zutshi}

\affiliation{Northern Illinois University, Dekalb, IL 60115, USA}

\begin{abstract}
We present two software packages developed to support detector R\&D
studies for the International Linear Collider.  LCDG4 is a
full-detector simulator that provides energy deposits from
particles traversing the sensitive volumes of the detector.  It has
been extensively used within the American ILC community, providing
data for algorithm development and detector optimization studies.
DigiSim models real-life digitization effects, converting the
idealized response into simulated detector readout.  It has many
useful features to improve the realism in modeling detector response.
The main characteristics of these two complementary packages are
discussed.
\end{abstract}

\maketitle

\thispagestyle{fancy}


\section{INTRODUCTION} 

Different detector technologies are currently under investigation to
be used for hadron calorimetry at the ILC (International Linear
Collider).  Each detector technology, with its own strengths and
weaknesses, may excel at different detector geometry setups.  The
challenging performance goal is the ability to perform at a
30\%/$\sqrt{E}$ jet energy resolution, which is a basic requirement
for resolving $Z$ from $W$ jets without ambiguity.  The {\it Particle
Flow Algorithm} (or PFA) is the most promising way to reach that goal.
Preliminary studies indicate that only joint optimizations of detector
technologies, geometries and reconstruction algorithms can satisfy
this jet energy resolution goal.

The Northern Illinois Center for Accelerator and Detector Development
(NICADD) at the Northern Illinois University (NIU) has been
developing, in collaboration with other software development groups,
basic software tools for the combined optimization of detector
technologies, geometries and reconstruction algorithms.  In this paper
we describe software packages which have been developed as tools for
detector optimization.  The NIU work on PFA algorithms is described
elsewhere\cite{dhiman-ref}.

\section{DETECTOR SIMULATION} 

In order to be useful for the detector optimization needs, a full
detector simulation package must be based on a simple and flexible
geometry description. It should allow all candidate technologies and
geometries to be tested, and provide the data for the development of
reconstruction algorithms.  An important point is that not only
simulation experts, but any interested user should be able to make
geometry changes to an existing detector description and analyze the
effects of those changes.

\subsection{The LCDG4 package}

LCDG4\cite{lcdg4-ref} is a general purpose, Geant4\cite{geant4-ref}-%
based full-detector simulation package.  The detector geometry is
described using a powerful and easy to change XML-based schema (called
{\it lcdparm}).  It is compatible with the I/O data formats used in
US, namely {\it STDHEP}\cite{stdhep-ref} for input, and {\it LCIO} or
{\it SIO} for output.  LCIO\cite{lcio-ref} has been adopted as the
standard data format to be used within the worldwide ILC community,
while SIO\cite{sio-ref} has been, and still is, extensively used
within the US ILC community.

The main limitation of LCDG4 is the limited number of geometrical
volumes available in the lcdparm schema used, allowing only disks,
cylinders and cones.  This limitation is not a problem when studying
global geometry parameters, but it is clear that the lcdparm schema
has to be extended to describe more realistic detectors.

\subsubsection{Geometry representation}

An example of the lcdparm XML schema for geometry representation is
illustrated in Fig.~\ref{lcdparm-f1}, which shows the lcdparm
description of a 55-layers hadronic calorimeter barrel.  Each layer
contains a 0.5 cm scintillator polystyrene layer preceded by 0.7 cm of
tungsten absorber.  The inner radius and the half-length of the barrel
cylinder are also shown.  The sensitive layers can be virtually
segmented, either projectively ({\tt segmentation} tag, for cells
subtending constant solid angle from the center of the detector) or
non-projectively ({\tt cell\_grid} tag, for cells of constant linear
dimensions).
\begin{figure*}[ht]
\centering
\fbox{ \ttfamily\tiny\bf
\begin{tabular}{l}
$<$?xml version="1.0"  ?$>$\\
$<$!DOCTYPE lcdparm SYSTEM "detParms.dtd" $>$\\
...\\
$<$lcdparm$>$\\
\ $<$global file="SDNPCalMar05-scint.xml" /$>$\\
\ $<$physical\_detector topology="silicon" id = "SDNPCalMar05-scint" $>$\\
...\\
\ \ $<$volume id="HAD\_BARREL" rad\_len\_cm="1.133" inter\_len\_cm="0.1193"$>$\\
\ \ \ \ $<$tube$>$\\
\ \ \ \ \ \ $<$barrel\_dimensions inner\_r = "138.26" outer\_z = "261.85" /$>$\\
\ \ \ \ \ \ $<$layering n="55"$>$\\
\ \ \ \ \ \ \ \ $<$slice material = "W" width = "0.7" /$>$\\
\ \ \ \ \ \ \ \ $<$slice material = "Polystyrene" width = "0.5" sensitive = "yes" /$>$\\
\ \ \ \ \ \ \ \ $<$slice material = "Air" width = "0.3" /$>$\\
\ \ \ \ \ \ $<$/layering$>$\\
\ \ \ \ \ \ $<$cell\_grid longit="1.0" transv="1.0" outer\_r="220.76"/$>$\\
\ \ \ \ \ \ $<$!--segmentation theta = "1024" phi = "1024" /--$>$\\
\ \ \ \ $<$/tube$>$\\
\ \ \ \ $<$calorimeter type = "had" /$>$\\
\ \ $<$/volume$>$\\
...\\
\  $<$/physical\_detector$>$\\
...\\
$<$/lcdparm$>$
\end{tabular}
} 
\caption{Example of a subdetector component described by the lcdparm
XML schema used in LCDG4.  Geometry changes are very easy to make.}
\label{lcdparm-f1}
\end{figure*}

\begin{figure*}[ht]
\centering
\includegraphics[height=46mm]{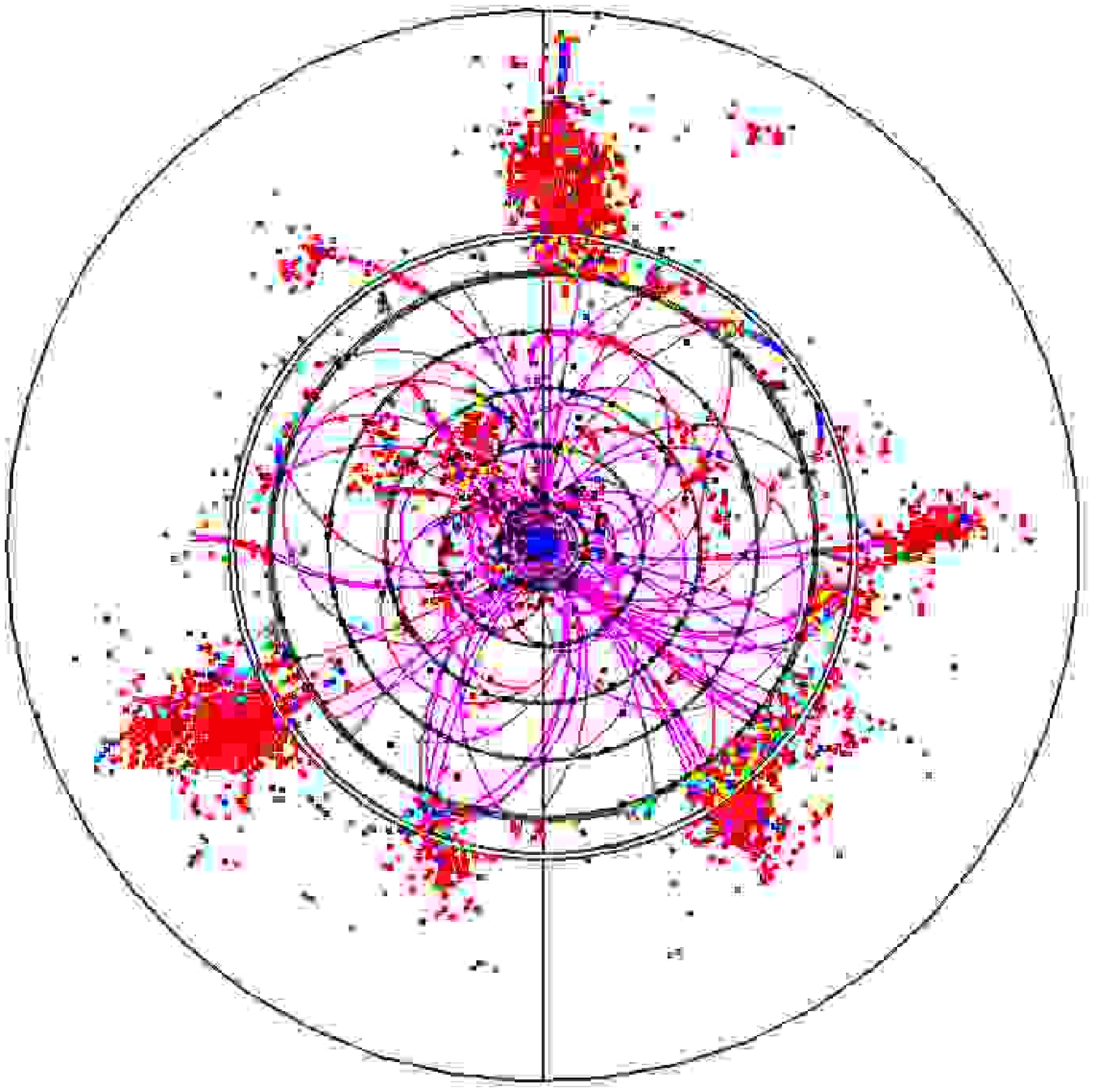}
\hspace*{1.cm}
\includegraphics[height=46mm]{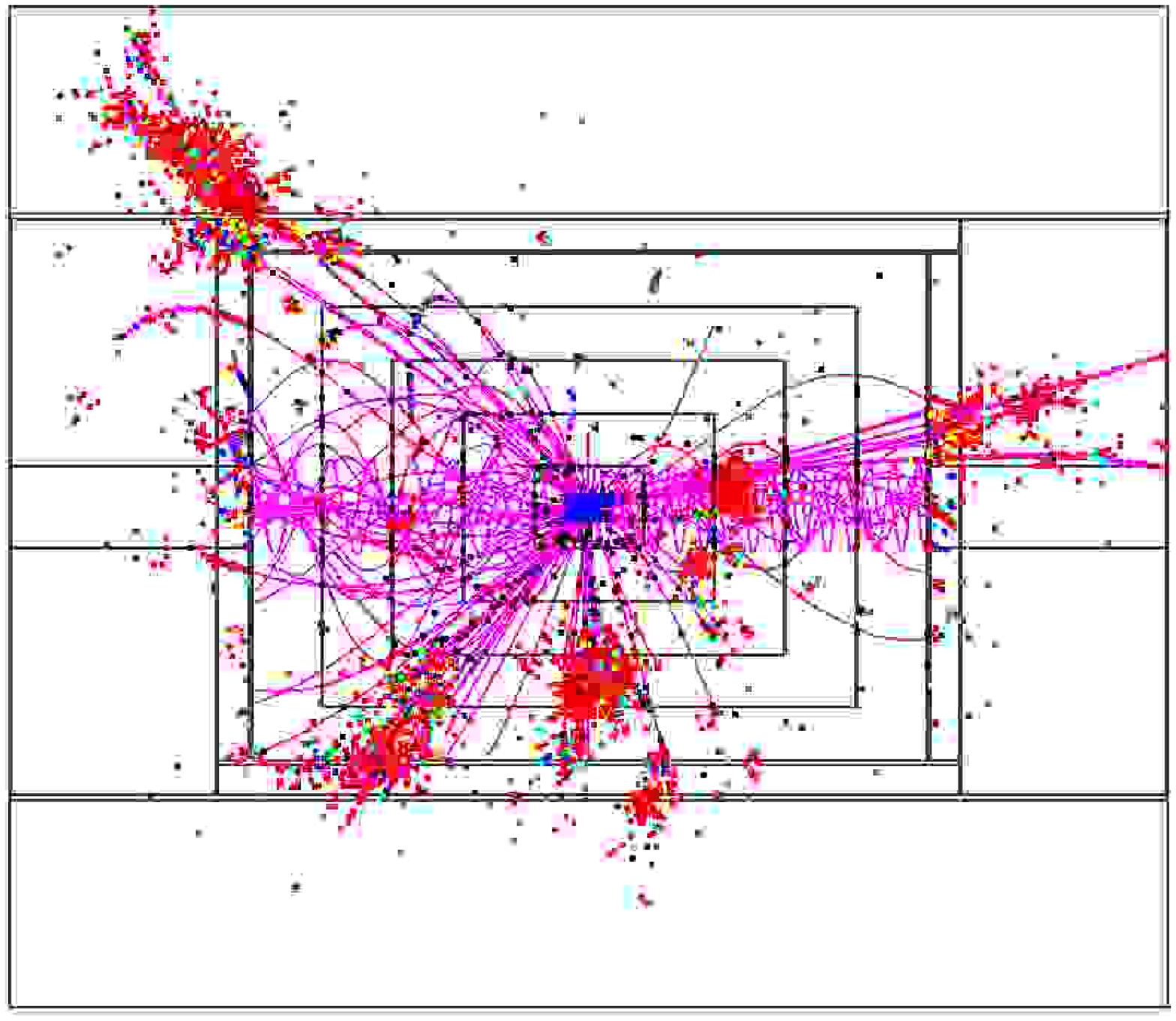}
\caption{Event display of a $t\bar{t}$ event, SDJan03 detector
geometry.  For clarity, neutral particles and the muon system
detectors are not shown, so all purple curves represent charged Monte
Carlo tracks.  Tracker hits are blue, while calorimeter hits, mostly
red, are in fact color-coded based on the hit energy.}
\label{evtdisp-f2}
\end{figure*}

Fig.~\ref{evtdisp-f2} shows the event display of a typical $e^+e^-
\!\!\rightarrow t\bar{t}$ event processed through LCDG4.  Many
trajectories can be easily followed in the highly segmented
calorimeters before showering starts.  Hits from individual particles
in jets can be separated using the full 3D information from the hits
in the jet shower, depending on the calorimeter granularity and
two-particle separation.

\subsubsection{Certification}

\begin{figure*}[ht]
\centering
\includegraphics[width=80mm]{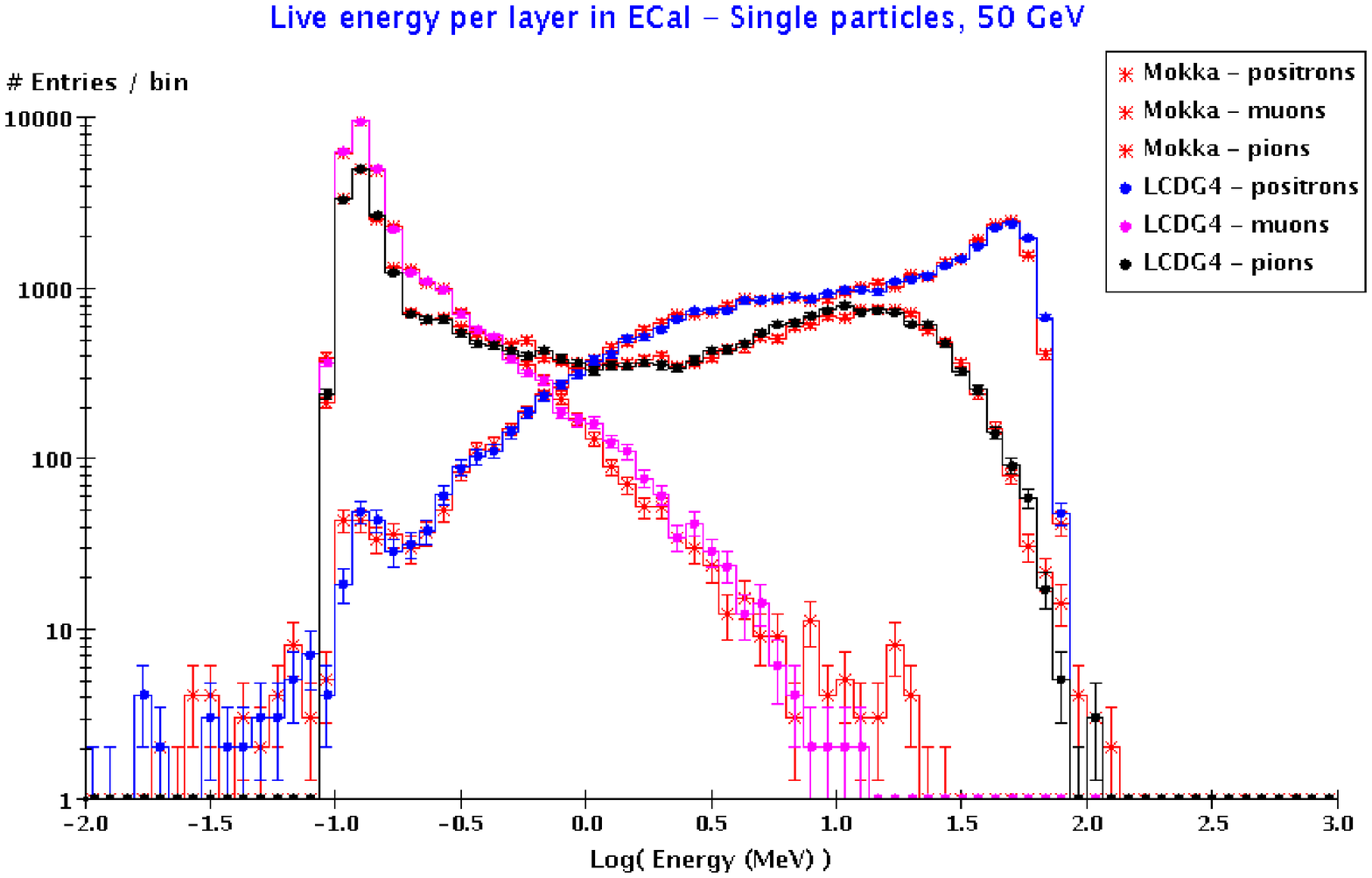}
\includegraphics[width=80mm]{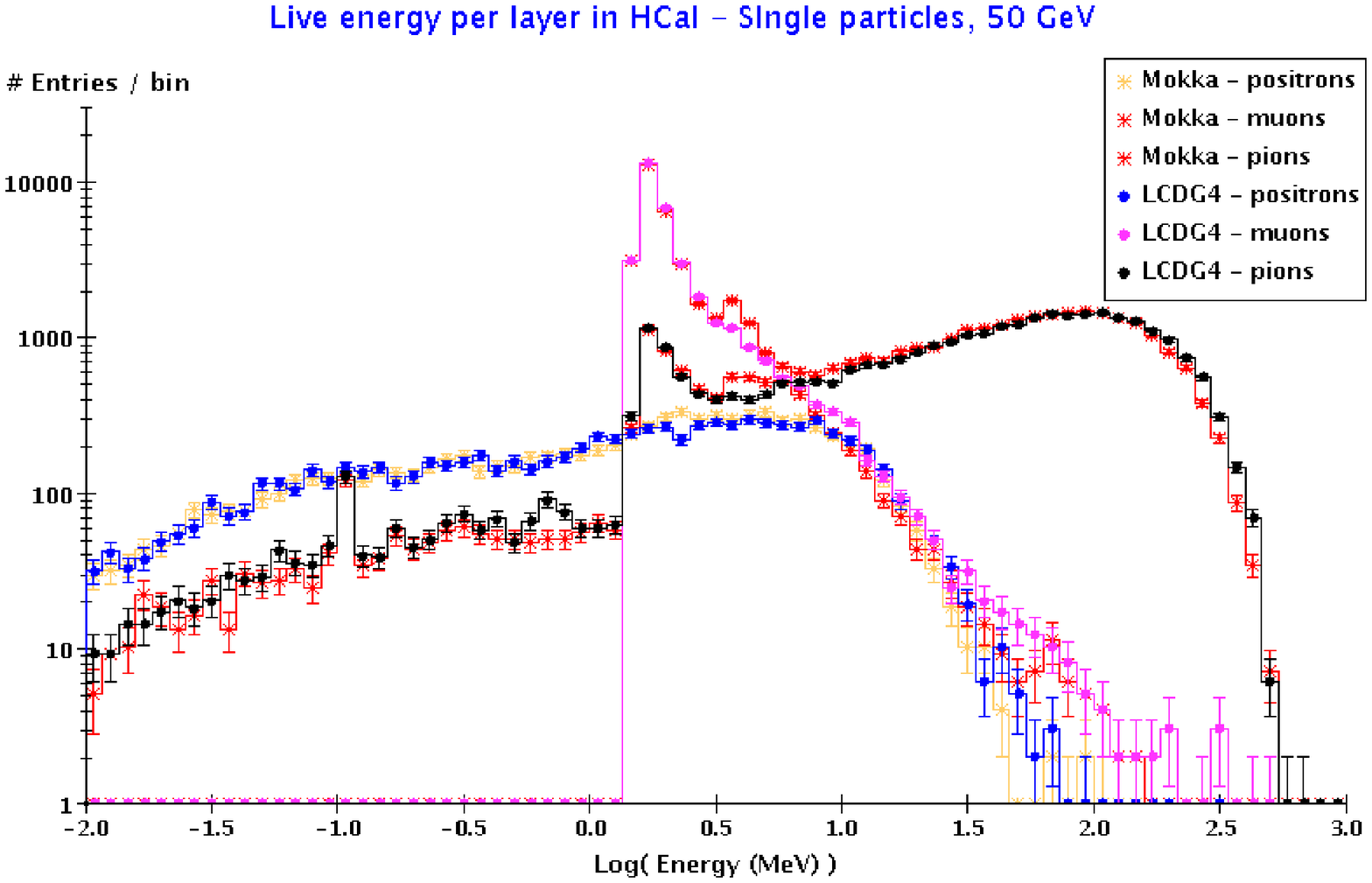}
\caption{LCDG4 data certification against Mokka v01-05: energy
deposition per layer in ECal (left) and HCal (right).}
\label{compMokka-f3}
\end{figure*}

The calorimeter simulation in LCDG4 was certified by comparing its
output with that of another detector simulator package,
Mokka\cite{mokka-ref}, which is mainly used in Europe.  We used the
then-current Mokka version 1.5.  Equivalent detector geometries were
implemented in both packages, and all relevant simulation parameters
were checked to be identical.  Material densities, radiation lengths,
etc.\ were equivalent to better than 0.1\%.  The distributions
compared include energy depositions per cell and per layer, hit
multiplicity, and longitudinal profile of energy distribution.  A
typical result of such comparison is shown in Fig.~\ref{compMokka-f3}.
There is a very good agreement between the two independent packages.

\subsubsection{Perspectives}

Several single-particle and physics samples have been processed
through LCDG4, and are freely available\cite{samples-ref}, in both SIO
and LCIO data formats.  These samples have been used in several analyses
and algorithm development studies presented at this workshop.

As mentioned before, the main limitation of LCDG4 was the small number
of supported geometrical volumes in lcdparm schema.  However, some
other user requirements have been suggested for the new generation of
detector simulation software.  Instead of adding new features to
LCDG4, a more appropriate path would be to develop a new Geant4
application from the ground up, using the best features of LCDG4,
Mokka and other simulators.

The new package, called SLIC\cite{slic-ref}, is currently under
development, and is discussed elsewhere in this proceedings.  This
means that LCDG4 will not have any further significant developments.
Instead, it will be basically maintained, and used for data production
until SLIC becomes ready for production.

\section{DIGITIZATION SIMULATION}

Beyond the ideal detector simulation software described in the
previous section, the appropriate comparisons of different detector
technologies must also include real life digitization effects present
in the data collection process.  Typical examples include signal
integration, amplification, attenuation, saturation, discrimination,
crosstalk and noise characteristics and some other effects that may
differ significantly between different detector technologies.

It is also important to have a common software framework for the
digitization simulation, which is general enough to allow for the
parametrization of the dominant characteristics of different
technologies, yet simple enough for hardware developers to introduce
their own parametrizations at will.

It is desirable that this package can be used either as a stand-alone
processor after the detector simulation, or as an on-the-fly
preprocessor to reconstruction/analysis tasks.  As a stand-alone
processor, DigiSim produces persistent output files, with {\it
identical} structure as the real data to be eventually collected,
except that the simulation output also contains the Monte Carlo truth
information.  On the other hand, when used as an on-the-fly
preprocessor, DigiSim needs to be available within the same framework
used for the reconstruction or analysis task.  Note that different
frameworks are commonly used in Europe (C++ based Marlin) and US (Java
based org.lcsim), so both of these languages are supported.

\subsection{The DigiSim package}

The purpose of the DigiSim package\cite{digisim-ref} is to do detector
digitization, for the CALICE\cite{calice-ref} beam tests as a first
goal, and ultimately for the full ILC detector(s).  It is readily
available for download and use, both as a standalone Marlin module (in
C++), and as one of the packages within the org.lcsim framework (in
Java).

The package reads the LCIO files produced by Geant4 applications (like
LCDG4, Mokka or SLIC) and appends the raw hits produced to the output
events. Most of the DigiSim (re)configuration can be done at run time
by editing an ASCII steering file, no recompilation is necessary.
Conveniently, the same ASCII file can be used for both the C++ and the
Java versions.

DigiSim has been designed to simulate the digitization process in
successive steps, performed by independent modules called {\it
modifiers}.  The existing modifiers (digitization classes) are easy to
setup and configure, and new functionality can be added easily by
creating new modifiers.  LCIO format is used for both input and output
files.  DigiSim is thus a powerful tool for detector technology
comparisons, very simple to use and extend.  It has been endorsed for
use within the CALICE collaboration, and is suitable for simulation of
calorimeter and tracker digitization for the ILC full-detector models.

\subsubsection{How DigiSim works}

\begin{figure*}[ht]
\centering
\includegraphics[width=100mm]{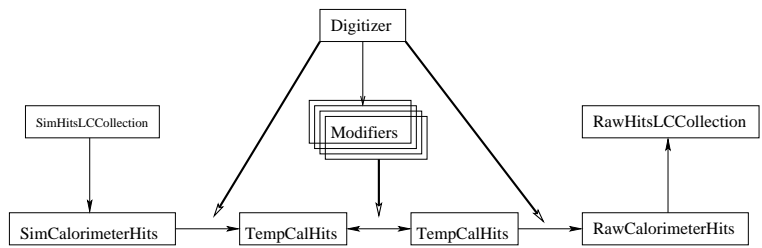}
\caption{DigiSim event loop.  Each {\tt Digitizer} class manages the
digitization processing for a given subdetector.  Transient {\tt
TempCalHit} objects are created from the simulated hits, and then
successively transformed by the full chain of modifiers.  The floating
point to integer conversion follows, and the resulting raw hits are
appended into the LCIO event structure.}
\label{digiLoop-f4}
\end{figure*}

The purpose of DigiSim is to simulate raw hits (integer ADC counts and
time stamps) starting from simulated hits (energy, timing and Monte
Carlo information in floating point variables).  DigiSim does this
using intermediate floating point objects ({\tt TempCalHit}), that
serve as both input to and output of the modifiers.  Ideally, each
modifier represents an individual digitization process, although some
very simple processes can be combined into a single modifier.  This
scheme is illustrated in Fig.~\ref{digiLoop-f4}.

In such a scheme, the full digitization process of a given subdetector
is represented by a chain of modifiers, managed by a digitizer (one
digitizer per subdetector).  Each modifier applies successive
transformations to their input hits, handing out their output hits to
the next modifier in the chain.  The output from the chain of
modifiers is used to create the raw hits, which are then simply
appended to the {\it LCEvent}.  The external framework takes care of
all the I/O involved.  Observe that the mapping between raw and
simulated hits is not one-to-one, due to crosstalk, noise and
threshold effects, but the final relationships between raw hits and
the original simulated hits is also stored into the DigiSim output.

\subsubsection{DigiSim implementation}

The DigiSim package was implemented originally in C++, to be used
within the CALICE collaboration\cite{calice-ref}, as a testbed for the
full detector digitization.  It was subsequently ported to Java, and
all developments since then have been performed in the Java version.
Consequently, the Java version has more features at this time, but we
intend to support stable equivalent versions in C++ and Java in the
near future.

\begin{figure*}[ht]
\centering
\includegraphics[width=100mm]{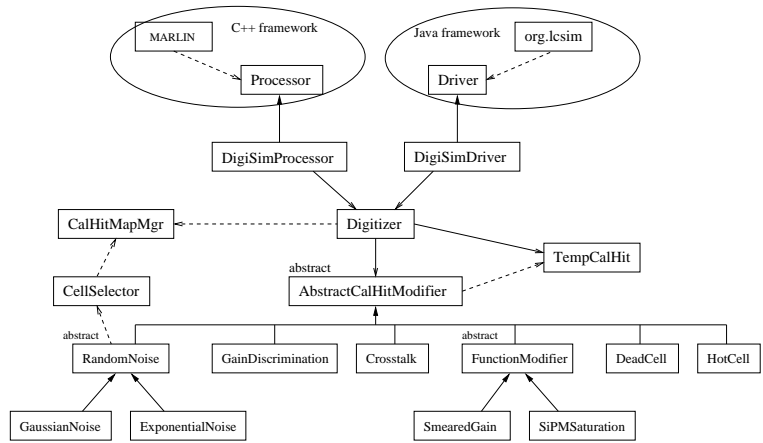}
\caption{DigiSim class diagram.  Full arrows represent
inheritance.  Hollow arrows represent containment (solid) or use 
(dashed) relationships.}
\label{digiClass-f5}
\end{figure*}

Fig.~\ref{digiClass-f5} shows a DigiSim class diagram, to explain how
the DigiSim implementations in Java and C++ use the same class names
and interfaces, as much as these languages allow.  The most
significant differences between the two implementations are in the
classes that interface DigiSim to their calling framework.  The names
of the interfacing classes make this clear: DigiSimProcessor inherit
from the Marlin Processor (C++), while DigiSimDriver inherit from the
org.lcsim Driver (Java).  These two classes adapt the common
interfaces of the core DigiSim classes to each corresponding
framework, simplifying considerably the process of porting any
development made in one programming language to another, thus
simplifying the synchronization between different implementations.

\subsubsection{Perspectives}

DigiSim documentation is available\cite{digisim-ref}, with detailed
instructions on downloading, installation, usage and also on how to
extend DigiSim, to enable the simulation of new digitization
processes, or modify any of the existing parametrizations or
algorithms.

The current implementation of DigiSim applies common parametrizations
to all the sensitive elements of a given subdetector.  Once the real
detector is built, DigiSim is expected to support different
cell-by-cell parametrizations.  Such configuration parameters may come
from a calibration database, for instance.

\section{STUDIES AND RESULTS}

\subsection{LCDG4}

Several ILC-related studies have been performed using simulation data
from LCDG4, particularly in the US.  Some of these studies have been
presented at this conference\cite{others-ref}.  LCDG4 has thus been
very useful for detector design and optimization studies, and several
new results are expected to be presented in the upcoming months,
including the Snowmass workshop (August 2005).

\subsection{DigiSim}

Most of the detector characteristics relevant to data acquisition and
digitization are still being studied by the detector hardware groups.
We plan to perform studies quantifying how the digitization of a
scintillator based hadronic calorimeter would affect the performance
of a typical PFA by comparing its performance on data before and
after DigiSim package processing.  For this purpose, we will need to
have dependable parametrizations for the relevant detector
digitization processes.  This is work in progress at NICADD/NIU.

For now, the results we can show about DigiSim just illustrate that
its output makes sense.  We emphasize that the current state of
parametrizations are only preliminary, often mere guesses.  With this
fact in mind, we present below some distributions to illustrate the
cumulative effect of each modifier along a given modifier chain.

As an example, we will show the effects of the following modifier
chain configuration, applied to the output of LCDG4.  The results
shown were obtained using the Java version of DigiSim, although
equivalent distributions should be obtained using the C++ version as
well, once the conversion to C++ is complete.

Fig.~\ref{digiConfig-f6} shows the DigiSim configuration file used, with
the chain of modifiers and their configuration parameters.  For
explanations or more details about the modifier parameters used, the
user is referred to the DigiSim documentation\cite{digisim-ref}.
\begin{figure*}[ht]
\centering
\fbox{ \ttfamily\tiny\bf
\begin{tabular}{l}
.begin HcalBarrDigitizer\\
\\
ProcessorType DigiSimProcessor\\
\\
InputCollection  HcalBarrHits\\
OutputCollection HcalBarrRawHits\\
Raw2SimLinksCollection HcalBarrRaw2sim\\
\\
ModifierNames HBlightYield HBcrosstalk HBlightCollEff SiPMQuEffic HBExpoNoise HBGaussNoise HBdiscrim HBGain \\
\\
\# Parameters:\\
\# modifierName    Type                gainNom  gainSig  thresh   thrSig\\
HBlightYield     GainDiscrimination  10000000        0       0        0\\
\\
\# Crosstalk                              mean    sigma\\
HBcrosstalk      Crosstalk               0.020     0.005\\
\\
\# Smeared gain parameters:          gain   gainSigma\\
HBlightCollEff   SmearedGain      0.0111   0.0029\\
SiPMQuEffic      SmearedGain        0.15        0\\
\\
\#\#\# Noise generators\\
\# GaussNoise parameters:        sys   be    Ecut   TimeNom  TSig   Mean  Sigma\\
HBGaussNoise   GaussianNoise      3     0      7       100   100    0.0  1.6\\
\# ExponentialNoise parameters:  sys    be   Ecut   TimeNom  TSig   Mean\\
HBExpoNoise    ExponentialNoise   3     0      7       100   100    0.6\\
\\
\# SiPM with saturation at about 2200 incident photons\\
\# SiPMSaturation parameters:          gainNom   linMax\\
HBSiPMSaturat    SiPMSaturation	            1     2200\\
\\
\# Hot cell parameters:      AmplNom  Sig  TimeNom  Sig      sys be lay the phi\\
HBHotCell       HotCell      252525   0    101010   0        3   0  12 123 345\\
\\
\# Hot cell parameters:          sys   be   lay   the   phi\\
HBDeadCell      DeadCell          3    0    12    34    56\\
\\
\# Discrimination                             threshold	 sigma\\
HBdiscrim	AbsValueDiscrimination   	   8       1\\
\\
\# Gain adjustments (temporary, test modifiers, etc)\\
HBGain          SmearedGain             60.06        0\\
\\
.end
\end{tabular}
} 
\caption{Example of a DigiSim configuration file for a scintillator-%
based barrel hadronic calorimeter.  Some of the parameters here model,
among other things, a fixed value of 10$^7$ photons per GeV of
deposited energy in the cell hit, with 2\% crosstalk to first
neighbors, followed by (1.11$\pm$0.29)\% efficiency for light
collection, 15\% quantum efficiency, exponential noise with 1.1
photoelectrons (PEs) in average, followed by a $1/4$ MIP threshold at
8 PEs.  Hot- and dead-cell modifiers are configured, but not really
used as they were not listed in the modifier chain.}
\label{digiConfig-f6}
\end{figure*}

Some explanation for the configuration file shown in
Fig.~\ref{digiConfig-f6} is in order.  The overall format of the file
is mandated by the Marlin framework, though by design, the same file
is also used for the Java version for convenience.  Lines starting
with \# are comments.  The first line is the name of the subdetector
digitizer.  After the LCIO collection names used (lines 3-5), names
for the modifiers making up the modifier chain for this subdetector
are listed in the order that they will be called.  Each of those
modifiers is then configured in the lines that follow.  For instance,
{\tt HBlightYield} is a {\tt GainDiscrimination}-type modifier that
requires 4 parameters.  First two parameters, say $g$ and $\sigma_g$,
are the factor and its smearing, $g\pm\sigma_g$, which are drawn for
each hit individually following a gaussian distribution -- the
so-called smeared linear transformation to hit energies.  The next two
parameters have a similar interpretation for a discrimination
threshold: the mean value and its smearing, also drawn for each hit
individually from a gaussian.  Note that we could have used a {\tt
SmearedGain} here as well, because no threshold was applied at this
level.

The modifier named {\tt Crosstalk} models light crosstalk at a rate of
2\% to the first neighbors, without second-order crosstalk.  Some
smearing has been used at the light collection modifier, namely
0.0111$\pm$0.0029 for a smeared value of 1.11\% for collection
efficiency.  Exponential and gaussian noise parametrizations are
available, and both are useful to model the noise of the photosensors,
although the noise numbers above are just guesses.  A 7~PE threshold
was used for both of these noise modifiers, affecting noise-only hits,
while a 8~PE cut was applied later by the {\tt HBdiscrim} modifier.

\begin{figure*}[ht]
\centering
\includegraphics[height=90mm]{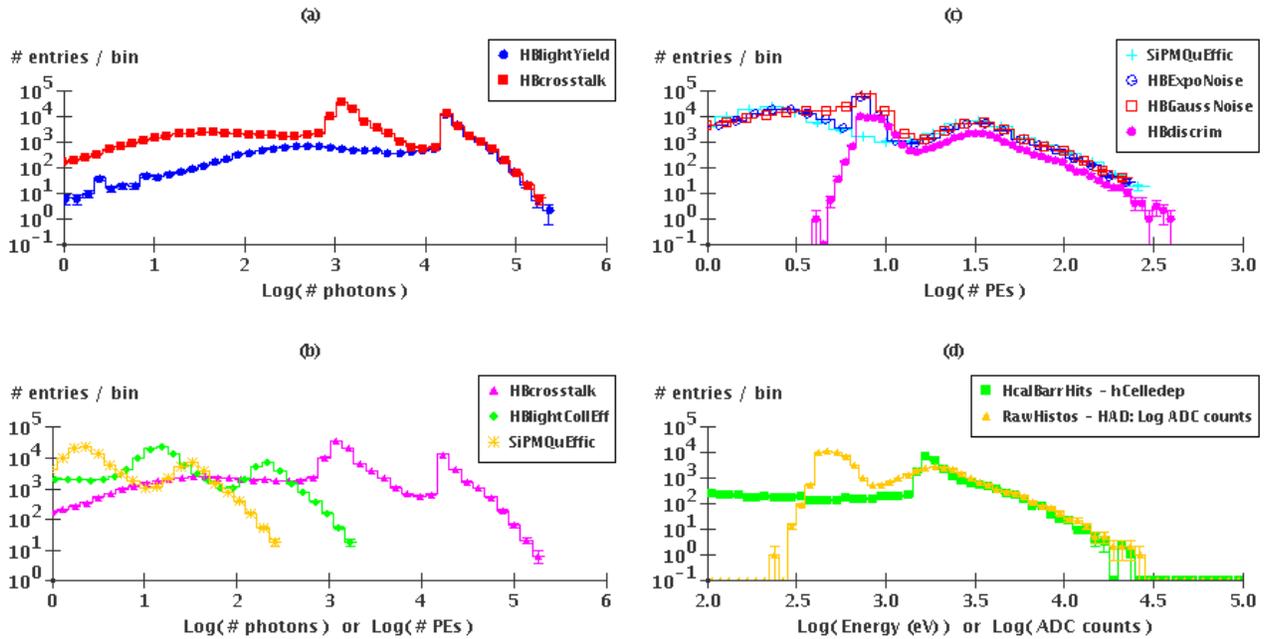}
\caption{Preliminary: Effects of the chain of modifiers configured as
in Fig.~\ref{digiConfig-f6} acting on the input energy distribution
for 500 10-GeV muons.  The x-axis should be interpreted differently
according to each modifier output: (a) dark blue shows the input
distribution, the energy deposition (from Geant4, in tenths eV) or,
equivalently, the number of scintillation photons.  Light crosstalk,
in \# photons, is shown in (a,red) and (b,pink).  Light collection
follows in (b,green), then photosensor quantum efficiency in \# PEs
(b,yellow and c,cyan), exponentially-distributed photosensor noise
(c,blue), gaussian electronics noise (c,red) and discrimination
(c,pink).  The last plot (d) compares the very input and output
distributions.  The gain used in the last step was chosen to bring the
MIP peak up to overlay the original distribution, for the sake of
comparison.}
\label{digiWorks-f7}
\end{figure*}

Fig.~\ref{digiWorks-f7} shows the result, in a MIP particles sample,
of the consecutive application of several modifiers, according to the
configuration file shown on Fig.~\ref{digiConfig-f6}. The smearing,
noise and crosstalk effects are apparent in the final (digitized)
distribution.

\section{SUMMARY AND PERSPECTIVES}

In this paper we have presented the current status of the
two most important NIU-developed tools currently in use, namely the
full detector simulation package LCDG4 and the digitization simulation
package DigiSim.

LCDG4 has been widely used within the US ILC community, specially for
those groups involved in calorimeter clustering and particle flow
algorithms.  Several important results based on LCDG4 have been
presented at this workshop.  LCDG4 will become deprecated, when SLIC
becomes ready for data production and algorithm developers stop using
SIO data in favor of the recommended LCIO.  Until then, we will
continue to maintain LCDG4, and use it to provide data for algorithm
development and detector optimization studies.

DigiSim, as a framework for digitization simulations, is ready for
production use.  Some extensions are foreseen, like independent
cell-by-cell configuration from a database.  It has been implemented
in both Java and C++ frameworks, using a common set of classes and
interfaces.

We presented the effect of a hypothetical configuration for
digitization simulation, comparing digitized (real) to simulated
(ideal) data, and the result of the incremental application of the
chain of modifiers.  The resulting plots demonstrate that DigiSim
package is producing results as expected.

\begin{acknowledgments}
The authors wish to thank the developers of org.lcsim and Marlin
frameworks.  This work is supported in part by grants from the US
Departments of Education and Energy, and the National Science
Foundation.
\end{acknowledgments}


\end{document}